# In-Plane Anisotropies of Polarized Raman Response and Electrical Conductivity in Layered Tin Selenide

*Xiaolong Xu, Qingjun Song, Haifeng Wang, Pan Li, Kun Zhang, Yilun Wang, Kai Yuan, Zichen Yang, Yu Ye,\* and Lun Dai\**

X. L. Xu, Q. J. Song, P. Li, K. Zhang, Y. L. Wang, K. Yuan, Z. C. Yang, Prof. Y. Ye, Prof. L. Dai
State Key Lab for Mesoscopic Physics and School of Physics, Peking University
Collaborative Innovation Center of Quantum Matter
Beijing 100871, China
E-mail: ye_yu@pku.edu.cn; lundai@pku.edu.cn
Dr. H. F. Wang
National Laboratory of Solid State Microstructures, School of Physics, Collaborative Innovation Center of Advanced Microstructures, Nanjing University
Nanjing 210093, China
Department of Physics, College of Science, Shihezi University
Xinjiang 832003, China

Abstract

The group IV-VI compound SnSe, with an orthorhombic lattice structure, has recently attracted particular interest due to its unexpectedly low thermal conductivity and high power factor, showing great promise for thermoelectric applications. SnSe displays intriguing anisotropic properties due to the puckered low-symmetry in-plane lattice structure. Low-dimensional materials have potential advantages in improving the efficiency of thermoelectric conversion, due to the increased power factor and decreased thermal conductivity. A complete study of the optical and electrical anisotropies of SnSe nanostructures is a necessary prerequisite in taking advantage of the material properties for high performance devices. Here, we synthesize the single crystal SnSe nanoplates (NPs) by chemical vapor deposition. The angular dependence of the polarized Raman spectra of SnSe NPs shows anomalous anisotropic light-mater interaction. The angle-resolved charge transport of the SnSe NPs expresses a strong anisotropic conductivity behavior. These studies elucidate the anisotropic interactions which will be of use for future ultrathin SnSe in electronic, thermoelectric and optoelectronic devices.

The group IV-VI compound, SnSe, with a puckered layered structure, has attracted extensive attention for a variety of potential applications due to its exotic physical properties. Bulk SnSe is a promising candidate for high-efficiency thermoelectrics due to its large Seebeck coefficient, high power factor, and low thermal conductivity. Along the b-axis of the crystal, the material has demonstrated an unprecedented ZT of 2.6 at 923 K.[1,2] Like other layered materials, SnSe is a promising two-dimensional (2D) semiconductor, with an indirect band gap of ~0.90 eV and a direct band gap of ~1.30 eV,[3] showing great potential for optoelectronic applications. Similarly, compared with their bulk counterparts, 2D nanostructures of these materials are expected to have a tunable band gap,[4, 5] arising from quantum confinement. Due to its unique puckered structure and weaker chemical bonds, the piezoelectric coefficient of single-layer SnSe was found to be one to two orders of magnitude larger than those of other 2D materials, such as $MoS_2$ and GaSe.[6] Moreover, single-layer SnSe is predicted to have in-plane ferroelectricity as it has a unique ionic-potential anharmonicity.[7] In addition, the reduced symmetry of this system leads to in-plane anisotropy which manifests in its various material properties. Recently, the phonon transport properties of single-layer SnSe have been systematically investigated, demonstrating the anisotropy of its lattice's thermal conductivity.[8] As many physical properties are dependent upon the phonon structure and lattice dynamics, Raman spectroscopy is an ideal method to elucidate the anisotropic light-matter interaction in these low-symmetry materials.[9-12] Moreover, studying the anisotropic charge transport behavior is necessary to optimize the device design and enhance the device performance. A complete study of the optical and electrical anisotropies of SnSe, which is still lacking so far, can enable its potential applications in novel electronic, thermoelectric and optoelectronic devices, where these anisotropic properties may become advantageous.

In this study, we successfully synthesized single crystal orthorhombic SnSe nanoplates (NPs) by chemical vapor deposition (CVD). The as-synthesized SnSe NPs were characterized by optical microscopy, atomic force microscopy (AFM), high-resolution transmission electron microscopy (HR-TEM), selected-area electron diffraction (SAED) and photoluminescence (PL) spectroscopy. We also systemically studied the optical and electrical anisotropies of the SnSe NPs by employing angle-dependent polarized Raman spectroscopy and angle-dependent conductivity measurements. The results show that the Raman of the SnSe NPs has a clear anisotropic dependence on polarization, photon, and phonon energies. We also observe a strong anisotropic electrical transport behavior with the conductivity along the zigzag direction larger than that of armchair direction, in good agreement with the theoretical prediction. These anisotropic studies provide us a deeper understanding of the materials' physical properties with respect to their structures, leading to the design of high performance electronic, thermoelectric, and optoelectronic devices.

The 2D SnSe NP was synthesized via CVD (**Figure 1a**, see Methods for detailed synthesis procedure).[13-16] The morphology of the as-synthesized SnSe NPs was characterized by optical microscopy and AFM. A typical optical image of the as-synthesized SnSe NPs (**Figure 1b**) shows that the SnSe NPs have rectangle shapes due to the orthorhombic structure. Each NP shows uniform optical contrast, indicating a uniform thickness across the whole plane of the NP, which is consistent with AFM measurements (**Figure 1c**). The AFM measurements display that the as-synthesized SnSe has a distribution from 10 nm to 210 nm in thickness. The thickness of the SnSe NPs can be controlled through the growth condition.

The HR-TEM image (**Figure 1e**) of a typical rectangular NP (inset of Figure 1e) shows clear orthogonal lattice fringes, with two similar lattice spacings of about 0.30 nm. The intersection angle between the lattice fringes is approximately 94°, which is in good agreement with the expected value for the (011) and ($0\bar{1}1$) planes of the orthorhombic SnSe crystal structure.[17, 18] The SAED pattern (**Figure 1f**) exhibits a clear orthogonally symmetry, indicating the single-crystal nature of the synthesized SnSe NP. We have measured more than five rectangular NPs and concluded that the top surfaces of the NPs belong to {100} crystal face, while the surfaces at the long side and short side of the rectangular NPs are {011} and {$0\bar{1}1$} crystal faces, respectively. The crystal growth follows the Gibbs-Curie-Wulff law. We performed first-principle calculations and found that the {011} planes with the lower surface energy are the preferable lateral face of the rectangular NP than {$0\bar{1}1$} face, because of its lower growth rate (See supporting information S1 for detail). The PL spectrum (**Figure 1d**) of the synthesized SnSe NP excited by a 633 nm laser shows a peak position around 1305 nm, in good agreement with the indirect bandgap of SnSe (0.9 eV).[19] All the results demonstrate that we have synthesized high quality single crystalline orthorhombic SnSe NPs providing a unique platform to study the anisotropic properties of the crystals.

Crystal structure of SnSe is shown in **Figure 2a-c**. It belongs to space group Pnma (62), with atoms arranged in two adjacent double layers of tin and selenium,[20] stacked along the a-axis through van der Waals interactions.[18] Herein, we define the slightly buckled zigzag direction (b-axis) as the y-axis, and the armchair direction (c-axis) as the z-axis, the direction perpendicular to the 2D plane (a-axis) as the x-axis. In the angle-dependent polarized Raman spectra measurement, the incident light was polarized along the vertical direction (**Figure 2d**) and an analyzer was placed before the entrance of the spectrometer, allowing for the analysis of the scattered Raman signals polarized parallel or perpendicular to the incident light polarization (called parallel- or cross-polarization configuration, respectively). The angular dependence of Raman response was obtained by rotating the sample in the y-z plane. The SnSe rectangle NP was initially placed arbitrarily. The angle θ (Figure 2d) is defined as the angle between the zigzag direction and laser polarization direction.

Series of Raman spectra of a SnSe NP excited by a 633 nm laser under parallel- (**Figure 2e**) and cross- (**Figure 2f**) polarization configurations with different sample rotation angles were measured in the 50-200 cm$^{-1}$ range. Altogether, four Raman modes can be resolved, which are $A_g^1$, $B_{3g}$, $A_g^2$ and $A_g^3$ modes peaked at 70, 106, 127 and 147 cm$^{-1}$, respectively. [9, 13] All of them can fit well using a Lorentzian lineshape. According to symmetry analysis, there should be 12 Raman-active modes of low symmetry SnSe ($4A_g+2B_{1g}+4B_{2g}+2B_{3g}$). Among them, only $A_g$ and $B_{3g}$ modes can be detected under our measurement configuration. Additionally, the resolution of our Raman system is limited so modes below 50 cm$^{-1}$ cannot be measured. Therefore, only four Raman modes ($A_g^1$, $B_{3g}$, $A_g^2$ and $A_g^3$ modes) can be resolved in our case. The Raman intensities are determined by Raman tensors (**Table 1**). Atomic displacements of these four modes are schematically plotted in **Figure 3**. The measured Raman frequencies agree well with those of the density functional theory (DFT) calculated phonon modes (Figure 3). All of these four Raman modes can be detected under the parallel-configuration, while the $B_{3g}$ mode disappears when the rotation angle is 0, 90 or 180 °(Figure 2e). The $A_g^2$ mode becomes very weak, while the $A_g^3$ becomes undetectable under the cross-configuration. The $B_{3g}$ mode disappears when the rotation angle approaches 60 or 150 °(Figure 2f). All the observations prove the polarized Raman spectra to be sensitive to the crystalline orientation of SnSe. For a quantitative study of the angular dependence of the Raman intensities, we summarized the polar plots for three typical Raman modes: $A_g^1$ (70 cm$^{-1}$), $B_{3g}$ (106 cm$^{-1}$) and $A_g^2$ (127 cm$^{-1}$) excited by 473 and 633 nm lasers (**Table 2**). The $A_g^3$ mode is undetectable except for the parallel-configuration under 633 nm laser excitation (see more details in **Figure S2 and S3**). We normalized the spectra by attributing the unity value to the most intense peak under the parallel-polarization configuration. The anisotropic Raman spectra of SnSe could be well understood by a classical Placzek model, where the Raman intensity of a phonon mode can be written as:[21]

$$I \propto |\vec{e}_i \cdot \tilde{R} \cdot \vec{e}_s|^2 \qquad (1)$$

in which $\vec{e}_i$ and $\vec{e}_s$ are the electric polarization unitary vectors of the incident and scattered lights, respectively, and $\tilde{R}$ is the Raman tensor for the Raman active modes of SnSe (Table 1). Based on the Cartesian coordinates denoted above, $\vec{e}_i$ and $\vec{e}_s$ are in the y-z plane. For an angle of θ, $\vec{e}_i = (0\ \cos\theta\ \sin\theta)$ for the incident light, while $\vec{e}_s = (0\ \cos\theta\ \sin\theta)$ and $\vec{e}_s = (0\ -\sin\theta\ \cos\theta)$ for the scattered light in the parallel- and cross-polarization configurations, respectively. As discussed above, a phonon mode can only be detected when $|\vec{e}_i \cdot \tilde{R} \cdot \vec{e}_s|^2$ has non-zero value. Therefore, using the above defined unitary vectors $\vec{e}_i$ and $\vec{e}_s$, as well as the Raman tensors of the $A_g$ and $B_{3g}$ modes, we can obtain the angular dependent intensity expressions for the $A_g$ and $B_{3g}$ modes to be:

$$S^{\parallel}_{A_g} = (|b|\cos^2\theta + |c|\cos\phi_{bc}\sin^2\theta)^2 + |c|^2\sin^2\phi_{bc}\sin^4\theta \qquad (2)$$

$$S^{\perp}_{A_g} = [(|b| - |c|\cos\phi_{bc})^2 + |c|^2\sin^2\phi_{bc}]\sin^2\theta\cos^2\theta \qquad (3)$$

$$S^{\parallel}_{B_{3g}} = (2|f|\sin\theta\cos\theta)^2 \qquad (4)$$

$$S^{\perp}_{B_{3g}} = [|f|\cos(2\theta)]^2 \qquad (5)$$

It is worth noting that, in an absorptive material, the elements of the Raman tensors should be complex numbers, with real and imaginary parts. In this case, the term $\phi_{bc}$ is the phase difference $\phi_b$-$\phi_c$, while the term $\phi_f$ is canceled out due to the square modulus of the Raman intensity expressions.

We can see that the angular dependencies of $A_g^1$, $B_{3g}$ and $A_g^2$ modes can be well described by equations 2-5 for the two different laser excitations (Table 2). In both parallel- and cross-polarization configurations, the $B_{3g}$ mode shows obvious 90 ° periodic variations with the sample rotation angle, yielding a 4-lobed shape with four maximum intensity angles under the different laser excitation. In parallel-polarization configuration, the four maximum intensity angles are 45°, 135°, 225° and 315° for $B_{3g}$ mode, while those are 0°, 90°, 180° and 270° in the cross-polarization configuration. The anisotropic Raman spectra of the $B_{3g}$ mode do not show clear excitation wavelength dependence. However, the anisotropic Raman spectra of the $A_g$ modes show obvious excitation wavelength dependences. In the parallel-polarization configuration under 473 nm laser excitation, $A_g$ modes show obvious 180° periodic variations, yielding a 2-lobed shape with two maximum intensity angles. However, under 633 nm laser excitation, the $A_g$ modes exhibit secondary maximum intensities at 90° and 270°, due to the large phase difference between the complex Raman tensor elements b and c. In the cross-polarization configuration under 473 nm laser excitation, $A_g$ modes show obvious 90 °periodic variations, yielding a 4-lobed shape with four maximum intensity angles at about 45°, 135°, 225° and 315°. Although the $A_g$ modes in the cross-polarization configuration under 633 nm laser excitation also exhibit 4-lobed shapes, the intensities at θ=45° and θ=225° are much smaller than those at θ=135° and θ=315°, which cannot be well fit by equation 3, because equation 3 predicts a 4-lobed shape with four equal maximum intensities under the cross-polarization configuration. In fact, we have measured several different samples under 633 nm laser excitation, and all of them turn out to be deformed 4-lobed shapes.

Similar abnormal phenomenon had also appeared in other anisotropic 2D material system.[11] However, it did not cause much attention. In this classical model, only two electron-radiation processes (absorption of the incident photon and emission of the scattered photon) are taken into account. Therefore, Raman scattering is based on the first order term in the Taylor expansion of the polarizability with respect to the vibrational normal coordinates.[22] In a full quantum mechanical model, the electron-phonon interaction is taken into account, which corresponds to a third-order process. Therefore, we believe that the measured anisotropies of $A_g$ mode in the cross-polarization configuration under 633 nm laser excitation, which is not in consensus with the classical model analysis, may be explained by involving the anisotropy of the electron-phonon interaction. So far, the electron-phonon interaction is still poorly understood, and it will be necessary for future work to fully elucidate the anisotropic properties of SnSe.

To get further insight of the anisotropic Raman response, we extract the ratio of the modulus in the Raman tensor elements and phase differences for the four detected modes into **Table 3**. In an anisotropic material such as SnSe, the incident photons with different energies will involve different electronic energy states upon the absorption. Accordingly, the |b|/|c| ratio and $\phi_{bc}$ would change with the excitation laser wavelength, which induces different anisotropies as the incident photon energy varies. This explains why the Raman anisotropies of the $A_g$ modes show obvious excitation laser wavelength dependence. On the other hand, for the $B_{3g}$ Raman mode, variation of the amplitude of the only Raman tensor element |f| does not change the polarization

dependence of the $B_{3g}$ mode. Therefore, the Raman anisotropy of $B_{3g}$ mode has no excitation wavelength dependence.

Recently, theory predicts the anisotropic electrical conductivity of SnSe, which is mainly due to the different effective masses of holes along the a, b, and c axes. As a result, the hole conductivity of SnSe along the zigzag direction should be larger than that of the armchair direction. We demonstrated that the SnSe NPs are *p*-type (see more details in **Figure S5**) by measuring the transport characteristics of the SnSe NPs through a field effect transistor. Then we performed the angle-resolved conductivity measurement of the SnSe NPs. 12 electrodes (220 nm Au) were fabricated on a SnSe NP spaced at an angle of 30° (**Figure 4a**). The width and the channel length of each pair of electrodes are 0.8 μm and 7 μm, respectively. We defined the zigzag direction as the 0° reference, which was confirmed by the angle resolved Raman spectra. The *I-V* curves were measured between each pair of electrodes separated at 180° apart (see details in **Figure S4a-b**). The obtained conductivity displays strong angle dependence (**Figure 4b**), showing a maximum conductivity along the zigzag direction and a secondary maximum conductivity along the armchair direction. This result is in agreement with the theoretical prediction,[23] with an experimentally opposite conductivity anisotropy compared to black phosphorus.[24] The 4-lobed shape of the conductivity anisotropy has not been observed in other 2D materials. We exclude the possibility of the contact resistance of each pair of electrodes through repetitive measurements on different devices (see more data in the **Figure S4c-d**). We ascribe the anisotropic electrical conductivity to the different effective masses of holes along different directions. Nevertheless, more detailed first principle calculations are needed to clarify this issue. The measured maximum conductivity ratio ($\sigma_{max}/\sigma_{min}$) is 3.9, which can be utilized to design anisotropic electronic/thermoelectric devices based on SnSe NPs. It is worth noting that the device geometry will spread the current and ultimately underestimate the conductivity anisotropy.

In summary, we have synthesized single crystal SnSe NPs using CVD. Polarized Raman responses of SnSe NPs under two laser excitation wavelengths (473 and 633 nm) were studied. The intricate dependences of the Raman anisotropy on polarization, photon, and phonon energies are observed and discussed in detail. The results reveal that both the anisotropic electron-photon and electron-phonon interactions contribute to an anisotropic Raman response of SnSe NPs. In addition, the electrical anisotropy of the SnSe NPs was studied using angle-resolved conductivity measurements. The result shows a maximum conductivity along the zigzag direction and a secondary maximum conductivity along the armchair direction, with a maximum conductivity ratio to be about 3.9. This work provides fundamental information for material/device designs, where the anisotropic properties might be utilized.

**Experimental Section**

*Synthesis of SnSe NPs*: SnSe NPs were grown inside a horizontal tube furnace with a 5 cm diameter quartz tube. SnSe powder (99.999%, Alfa Aesar) was placed at the center of the furnace in a quartz boat, and the freshly cleaved mica substrates were placed downstream at certain locations to maintain the deposition temperature in the range 350-450 ℃. The quartz tube was pumped and refilled with ultra-pure argon gas more than three times to reduce oxygen contamination, and then the furnace was heated to 650 °C in 17 min, and kept at this temperature for 5 min with a low growth pressure and an argon flow rate of 200 standard cubic centimeters per minute (sccm). After that, the furnace was cooled down to room temperature naturally.

*AFM and Raman Characterization*: The NPs' thickness and roughness was characterized by an AFM (Asylum Research Cypher S). Polarized Raman experiments were conducted using a confocal microscope spectrometer (Horiba JobinYvon HR800) with a 100× objective lens, and 473 nm (2.63 eV) and 633 nm (1.96 eV) lasers. The incident laser beam was polarized, and an analyzer was placed before the entrance of the spectrometer, allowing for the investigation of the scattered light polarization along directions parallel and perpendicular to the incident light polarization. The sample was rotated every 10° from 0° to 360°, yielding the measurement of the angular dependence of the Raman intensities in both polarization configurations. In every step, the microscope focus was optimized, and the position of the sample was adjusted to ensure the same point was detected.

*Angle-resolved conductivity measurement*: The SnSe NPs from mica substrate were transferred to a 300 nm $SiO_2$/Si substrate, where pre-patterned markers had been defined.[22] Then, PMMA (ethyl lactate, 950 kMW) of about 280 nm in thickness was spun on the substrate and baked at 170° for 5 min. Later, the star-shaped 12 electrodes (220 nm Au) were defined on a SnSe NP by electron-beam lithography (FEI NanoSEM), followed by development, thermal evaporation and lift-off processes. All electrical characterizations were performed using a Keithley 4200 semiconductor parameter analyzer.

**Acknowledgements**

This work was supported by the National Basic Research Program of China (Nos. 2013CB921901 and 2012CB932703), and the National Natural Science Foundation of China (Nos. 61521004, 61125402, 51172004, and 11474007). Y.Y. thanks M.Z. of the University of California, Berkeley for manuscript proofreading.

**Figures and Captions:**

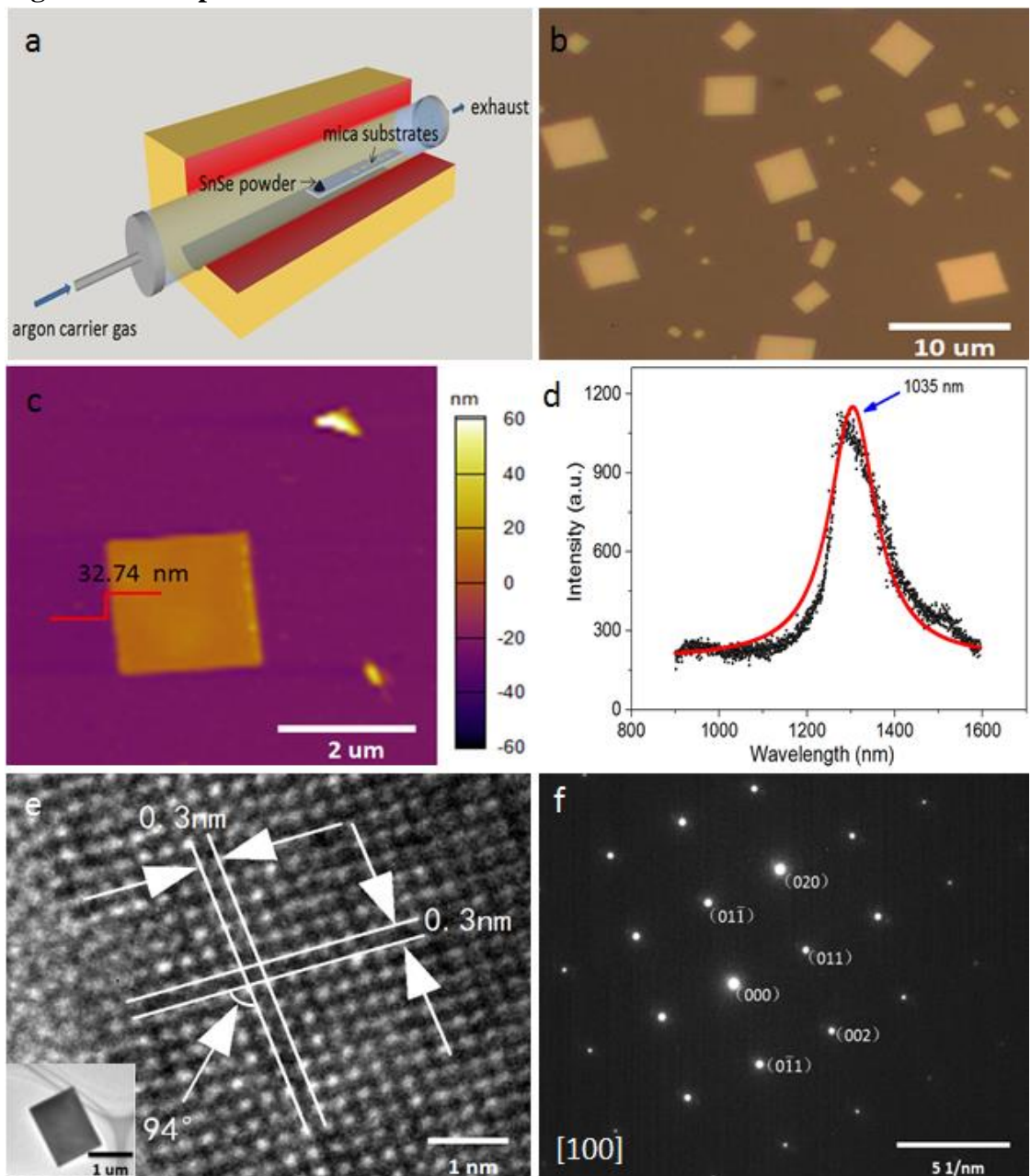

Figure 1. Optical microscopy, AFM, PL, and HR-TEM characterizations of the synthesized SnSe NPs. a) Schematic diagram of CVD synthesis of SnSe NPs. The quartz tube was filled with argon as carrier gas and pumped to keep a low growth pressure. b) Optical image of the synthesized SnSe NPs on the mica substrates, showing a size distribution of 2 μm to 8 μm. c) AFM image of a typical SnSe NP, showing a uniform thickness of 32.7 nm. d) PL spectrum of the SnSe NPs shows a peak intensity at 1305 nm, which is in good agreement with the reported indirect bandgap of SnSe (0.9 eV). e) HR-TEM image of a SnSe NP. Inset: the low-magnification TEM image of the SnSe NP. f) The corresponding SAED pattern measured along the zone axis [100], confirming the single crystalline orthorhombic nature. Statistic TEM studies prove the long sides of the rectangular NPs belong to {011} crystal face and the short sides belong to {0$\bar{1}$1} crystal face.

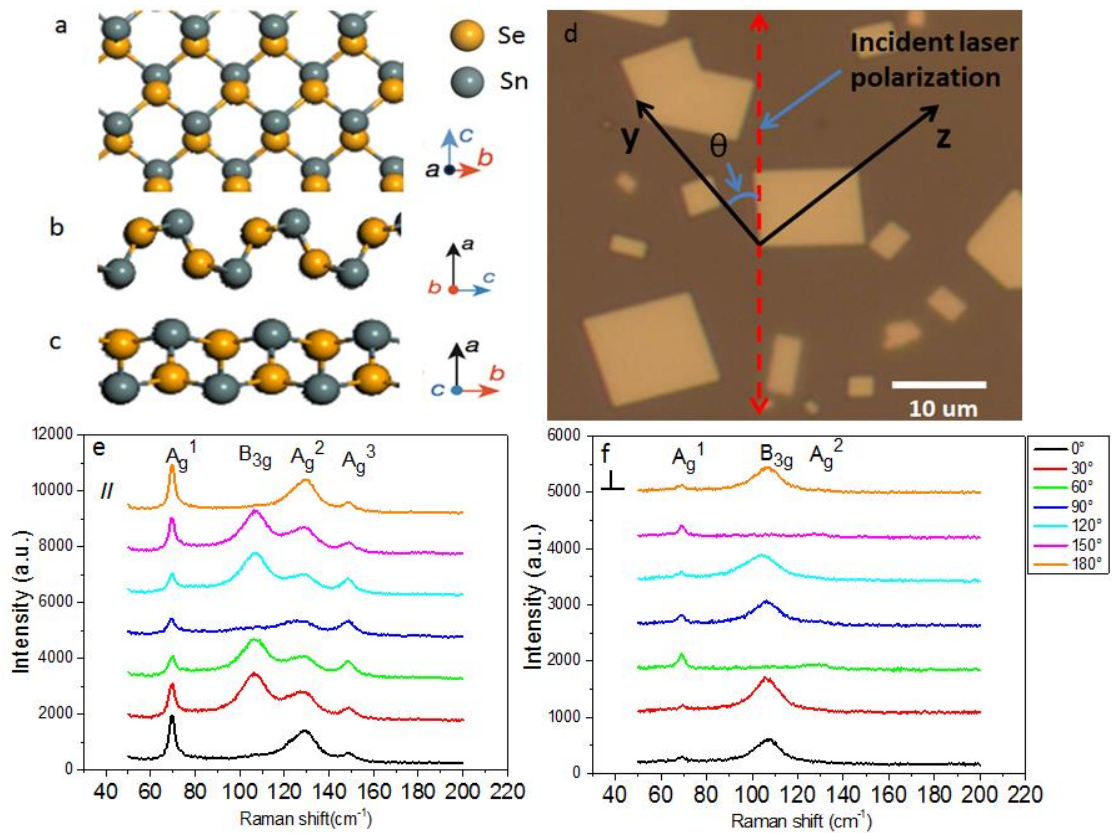

Figure 2. Crystal structrue and polarizaed Raman spectra of the SnSe NPs. a) top view, b) front view and c) side view cystal structures of single-layer SnSe. The armchair and zigzag directions are along the c axis and b axis, respectively. d) Optical image of a measured NP. The crystallographic axes, the incident light polarization direction, and the angle θ between the incident light and the y axis were indicated. e)-f) Typical polarized Raman spectra of a SnSe NP with different sample rotation angles excited by a 633 nm laser under the parallel-polarization configuration (e) and cross-polarization configuration (f), showing obvious angle dependences. Note that the $A_g^2$ mode in the cross configuration is much weaker compared with other modes but it is still strong enough to be well fitted by Lorentzian lineshape.

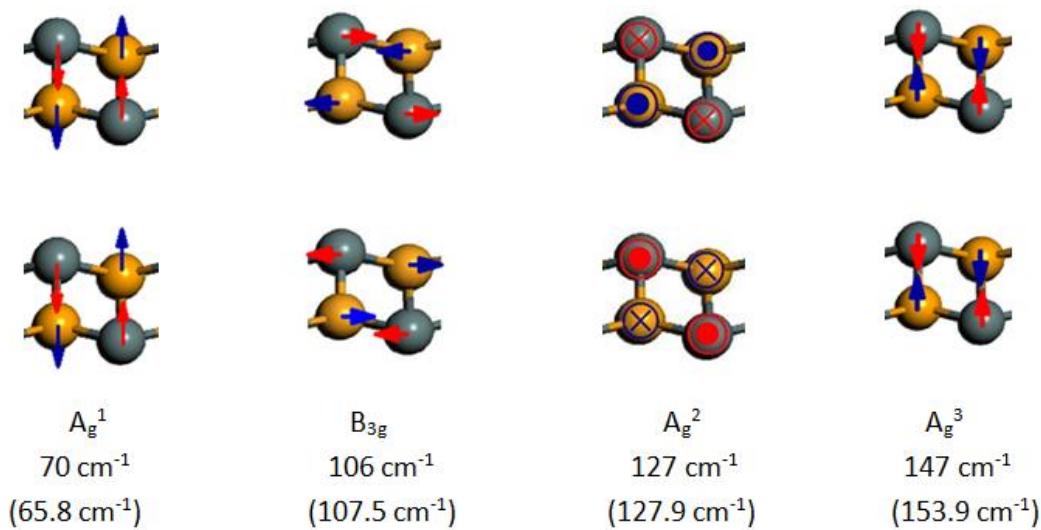

Figure 3. Atomic displacements of the four modes measured in the 50-200 cm$^{-1}$ range. The calculated frequencies are provided in parenthesis, which are in good agreement with experimental values.

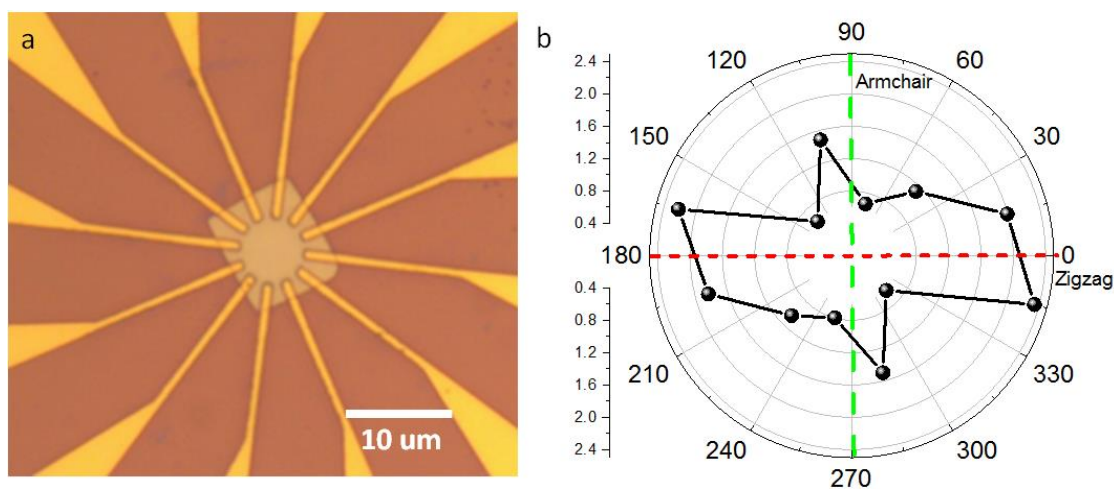

Figure 4. Angle-resolved conductivity of SnSe NPs. a) Optical image of SnSe NP with 12 electrodes spaced at 30° apart. The width and the channel length of each pair of electrodes are 0.8 μm and 7 μm, respectively. b) The polar plot of the angle-resolved conductivity. The crystalline orientation was determined by polarized Raman spectra measurements. The angle-dependence conductivity displays strong anisotropy, with a maximum conductivity along the zigzag direction and a secondary maximum conductivity along the armchair direction.

**Tables and Captions:**

| mode | $A_g$ | $B_{1g}$ | $B_{2g}$ | $B_{3g}$ |
|---|---|---|---|---|
| Raman tensor | $\begin{bmatrix} \|a\|e^{i\phi_a} & 0 & 0 \\ 0 & \|b\|e^{i\phi_b} & 0 \\ 0 & 0 & \|c\|e^{i\phi_c} \end{bmatrix}$ | $\begin{bmatrix} 0 & \|d\|e^{i\phi_d} & 0 \\ \|d\|e^{i\phi_d} & 0 & 0 \\ 0 & 0 & 0 \end{bmatrix}$ | $\begin{bmatrix} 0 & 0 & \|e\|e^{i\phi_e} \\ 0 & 0 & 0 \\ \|e\|e^{i\phi_e} & 0 & 0 \end{bmatrix}$ | $\begin{bmatrix} 0 & 0 & 0 \\ 0 & 0 & \|f\|e^{i\phi_f} \\ 0 & \|f\|e^{i\phi_f} & 0 \end{bmatrix}$ |

Table 1．Raman tensor forms for all active modes in SnSe.

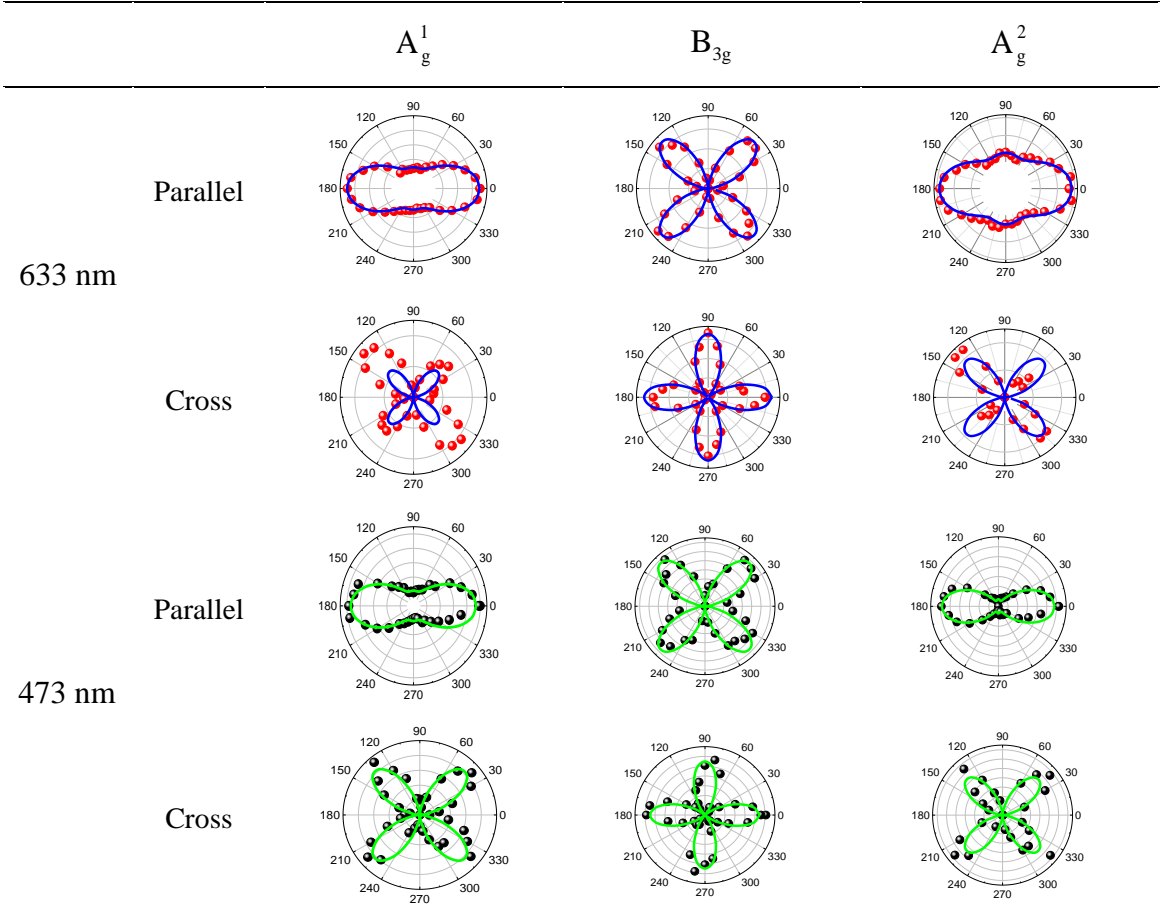

Table 2. Angular dependence of the Raman intensities measured using the 633 and 473 nm lasers under both parallel- and cross-polarization configurations. The solid dots are experimental data, while the solid curves correspond to the best fitting to the data using Equations 2-5 for each configuration and mode symmetry. Note that in all the polar figures the horizontal direction is along the armchair direction and the vertical direction is along the zigzag direction.

| Experimental frequency (cm$^{-1}$) | | 70 | 106 | 127 | 147 |
|---|---|---|---|---|---|
| Irreducible representation | | $A_g^1$ | $B_{3g}$ | $A_g^2$ | $A_g^3$ |
| Calculated frequency (cm$^{-1}$) | | 65.8 | 107.5 | 127.9 | 153.9 |
| Raman tensor elements $\|b\|, \|c\| and \|f\|$ | 633 nm | $\frac{\|b_1\|}{\|c_1\|} = 1.76$ | $\|f\| = 0.99$ | $\frac{\|b_3\|}{\|c_3\|} = 1.41$ | $\frac{\|b_4\|}{\|c_4\|} = 0.76$ |
| | 473 nm | $\frac{\|b_1\|}{\|c_1\|} = 2.94$ | $\|f\| = 0.58$ | $\frac{\|b_3\|}{\|c_3\|} = 3.18$ | |
| $\Phi$ (°) | 633 nm | 61 | 0 | 60 | 17 |
| | 473 nm | 69 | 0 | 0 | |

Table 3．The experimental frequencies, irreducible representation, calculated frequencies, Raman tensor elements, and phase difference for all the detected phonon modes.

**In-Plane Anisotropies of Polarized Raman Response and Electrical Conductivity in Layered Tin Selenide**

*Xiaolong Xu, Qingjun Song, Haifeng Wang, Pan Li, Kun Zhang, Yilun Wang, Kai Yuan, Zichen Yang, Yu Ye,\* and Lun Dai\**

**S1. The surface energy calculation**

For the as-synthesized rectangular SnSe NPs, our TEM studies prove that the surfaces at the long side and short side are {011} and {0$\bar{1}$1} crystal faces, respectively. This phenomenon could be understood by Gibbs-Curie-Wuff law, stating that the normal growth rates of crystal faces are proportional to the corresponding surface free energies. We calculated the energies of SnSe (011) and (0$\bar{1}$1) surfaces by the first-principles calculations based on density functional theory (DFT) and the generalized gradient approximation (GGA), using the Vienna ab initio simulation package (VASP). The surface energy is defined by the follow function:

$$\varepsilon = \frac{E(nSnSe) - nE(SnSe)}{2A}$$

where $E$(nSnSe) is the total energy of SnSe surface slab. n is the number of atoms in the surface slab. $E$(SnSe) is the bulk energy per atom. $A$ is the surface area. SnSe surfaces (011) and (0$\bar{1}$1) were modeled using periodic slabs with ten SnSe bilayers separated by a 10 Å vacuum region to yield and surface energies are 0.026 and 0.029 eV/Å$^2$, respectively. According to the Gibbs-Curie-Wulff law, the (011) surface with the lower surface energy is the preferable lateral face of SnSe crystal rather than (0$\bar{1}$1) surface, because of its lower growth rate. The surface energies (002)> (0$\bar{1}$1)> (011), indicating the growth rate (002)> (0$\bar{1}$1)> (011). At some growth nodes, a special shape with truncated corners will appear in the as-synthesized SnSe NPs (**Figure S1a-b**), which is consistent with our calculation. With the continuous growth, the truncated corners will be filled (**Figure S1c**).

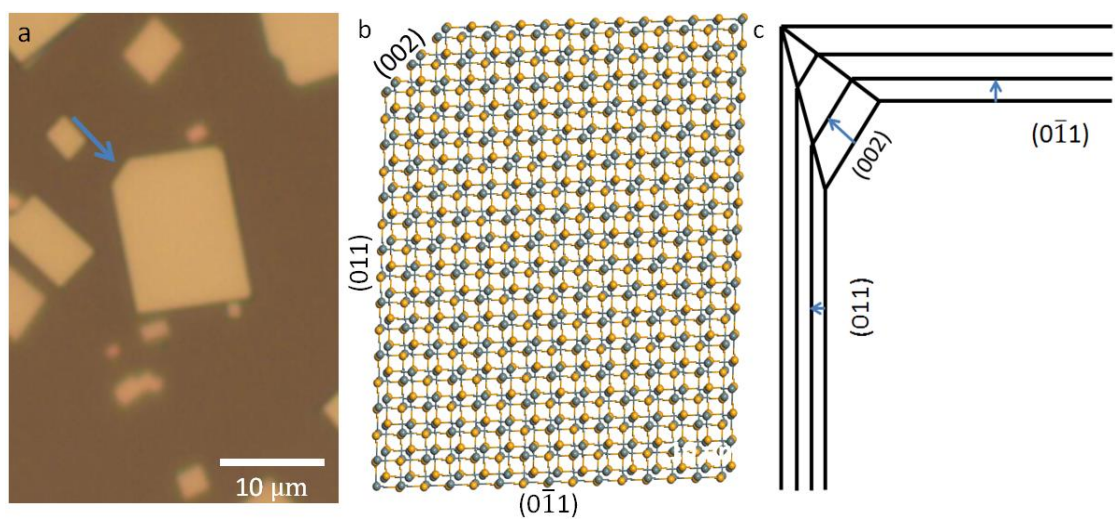

Figure S1. a) Optical micrograph of as-synthesized SnSe NP on the mica substrates. A typical NP shows a shape with a truncated corner. b) Top view ball-and-stick schematic illustration of corresponding SnSe NP in (a). The corresponding surfaces are labeled. c) Schematic diagram of the single crystal NP growth process. According to the Gibbs-Curie-Wulff law, the truncated corner will eventually be filled because of the different growth rates of the different surfaces.

**S2. The angle-dependent Raman intensity spectra**

The angle-dependent Raman intensity spectra of SnSe NPs under 633 nm (**Figure S2a-b**) and 473 nm (**Figure S2c-d**) laser excitations are measured by rotating the sample in the parallel- and cross-polarization configurations. We can clearly see that different Raman modes display different angle dependences. The $A_g^3$ mode is undetectable except for the parallel-configuration under 633 nm laser excitation.

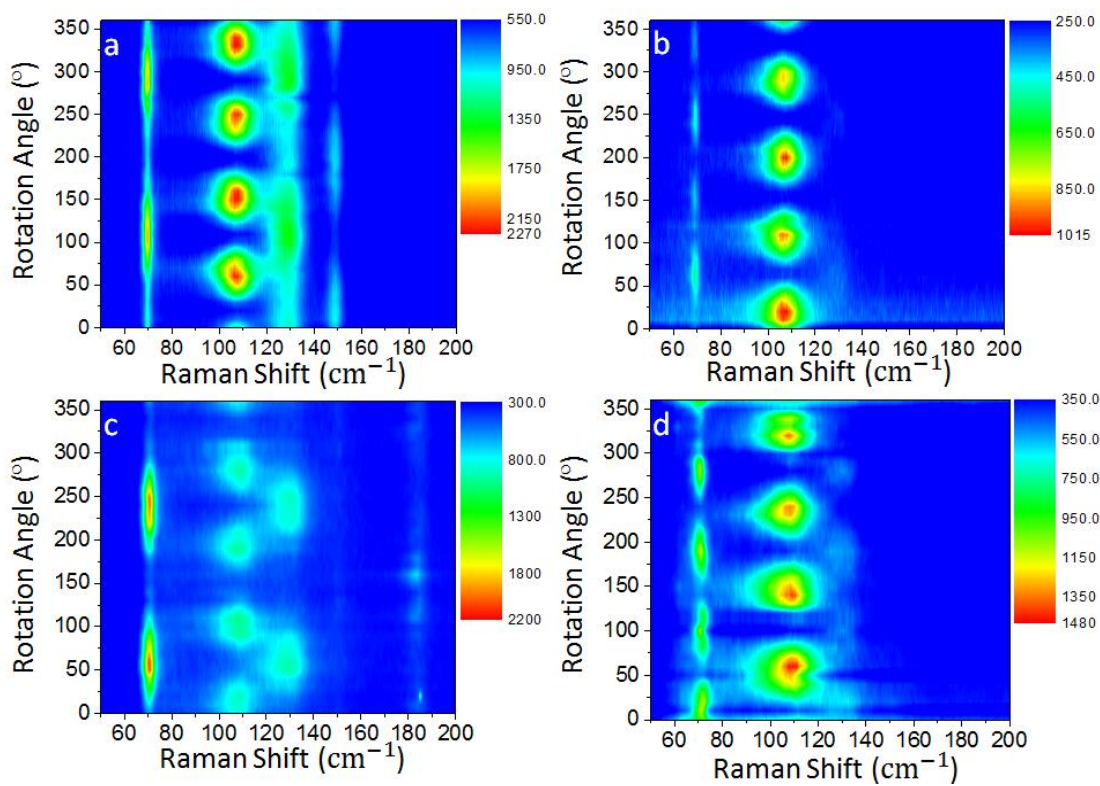

Figure S2. Angle-dependent Raman intensity spectra of SnSe NPs. a)-b) Raman spectra under 633 nm laser excitation in parallel- and cross-polarization configuration, respectively. c)-d) Raman spectra under 473 nm laser excitation in parallel- and cross-polarization configuration, respectively.

**S3. Complete polar plots of Raman modes under 633 nm laser excitation**

The $A_g^3$ mode is undetectable except for the parallel-configuration under 633 nm laser excitation. The complete polar plots of the Raman intensities under 633 nm laser excitation in both parallel- and cross-polarization configurations (**Figure S3**) show how different Raman modes have different angle dependences. The $A_g^3$ mode in parallel-configuration also shows obvious 180° periodic variations, yielding a 2-lobed shape with two maximum intensity angles at 90° and 270°.

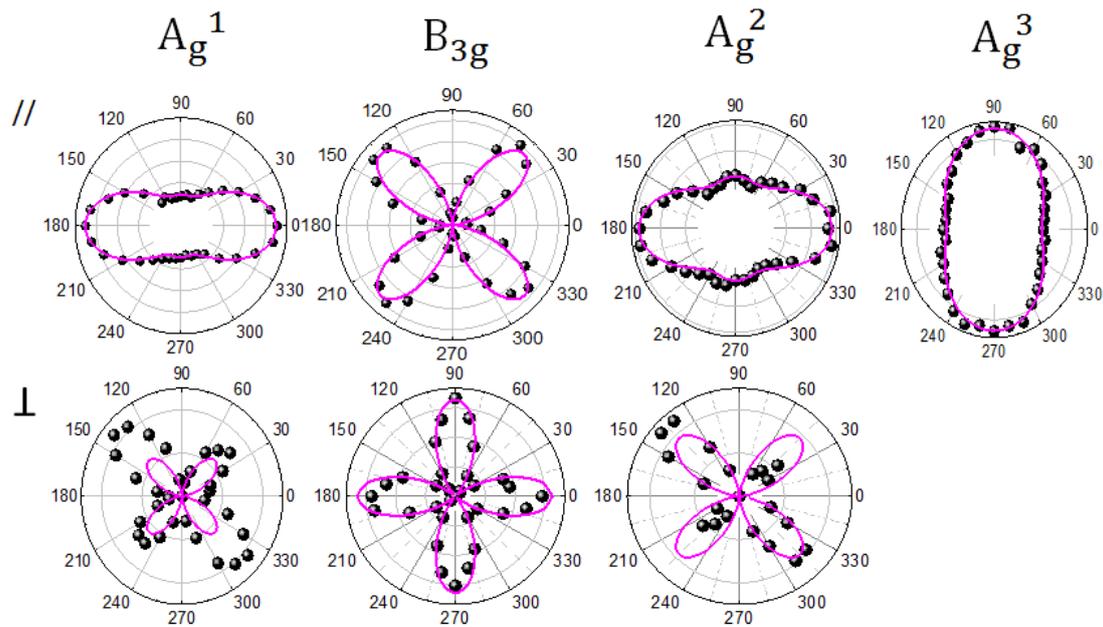

Figure S3. Angular dependence of the Raman intensities measured under 633 nm laser excitation in both parallel- and cross-polarization configurations. Dots are the experimental data, while the solid curves correspond to the best fitting to the data using Equations 2-5 in main manuscript for each configuration.

**S4. Angle-resolved conductivities of SnSe NPs**

The anisotropic conductivity was measured by *I-V* curves across each pair of diagonally positioned electrodes separated at 180° apart. The *I-V* curves of two different devices (**Figure S4a-b**) both show obvious angle dependences. The polar plots of the angle-resolved conductivity of these two devices share similar shapes (**Figure S4c-d**). The crystalline orientations were determined by polarized Raman spectra measurements.

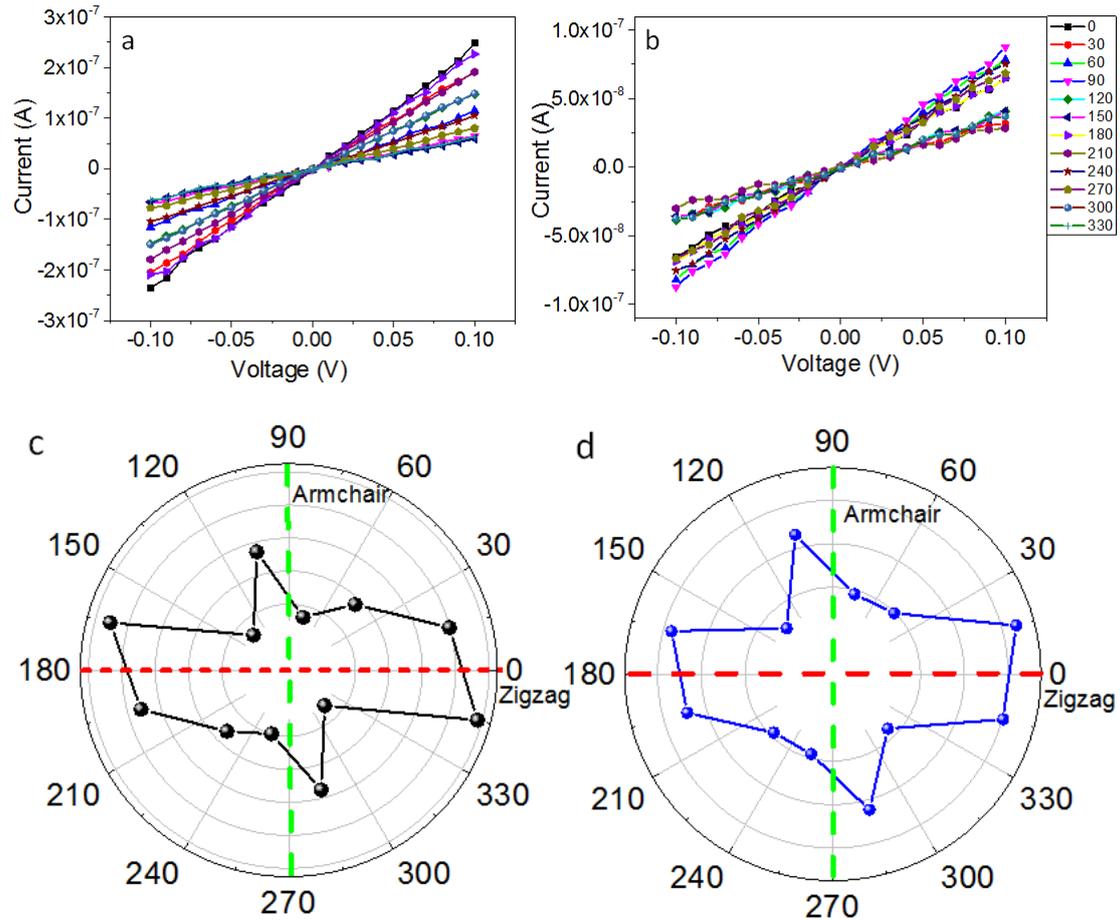

Figure S4. Angle-resolved conductivities of SnSe NPs. a)-b) The *I-V* curves of different pairs of diagonally positioned electrodes in two different devices with bias voltages from −0.1V-0.1 V. c)- d) The polar plots of the angle-resolved conductivities of these two devices. The crystalline orientations were determined by the polarized Raman spectra.

**S5. Transport characterization of the SnSe NP**

We measured the transport characterization of the SnSe NP in a transistor structure. The drain current decreases with the increase of gate voltage, showing that the SnSe NPs are hole doped (**Figure S5**).

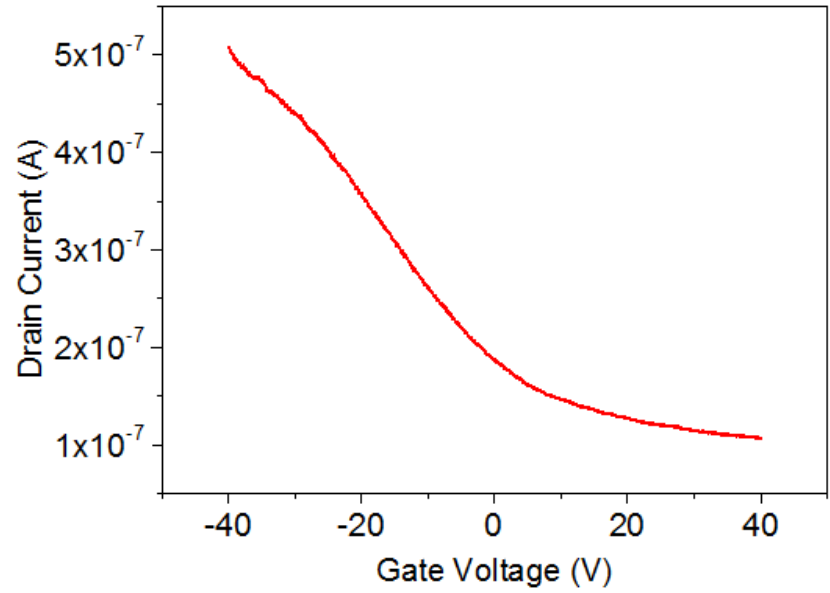

Figure S5. Transfer curve of the SnSe NP transistor at $V_{ds} = 1$ V, indicating that the SnSe NPs are hole doped.